\documentclass[reprint, superscriptaddress, secnumarabic, amssymb, nobibnotes, aps, prl]{revtex4-2}

\usepackage{graphicx}
\usepackage{epstopdf}
\usepackage[T1]{fontenc}
\usepackage[latin9]{inputenc}
\usepackage{amsbsy}
\usepackage{gensymb}
\usepackage{textcomp}
\usepackage[T1]{fontenc}
\usepackage[latin9]{inputenc}
\usepackage{amsmath}
\usepackage{amssymb}
\usepackage{bbm}
\usepackage{soul}
\usepackage{physics}
\usepackage{braket}
\usepackage{xcolor}
\allowdisplaybreaks
\usepackage{graphicx}
\usepackage[colorlinks=true]{hyperref}  
\hypersetup{
    unicode=false,          
    pdftoolbar=true,        
    pdfmenubar=true,        
    pdffitwindow=false,     
    pdfstartview={FitH},    
    pdftitle={Superconducting Ground State of Cr-Based Equiatomic High-Entropy alloy through \texorpdfstring{\(\mu\)SR}{muSR}},    
    pdfauthor={, , ,},     
    pdfsubject={},   
    pdfcreator={},   
    pdfproducer={},  
    pdfkeywords={} {} {}, 
    pdfnewwindow=true,      
    colorlinks=true,       
    linkcolor=blue, 
    citecolor=blue,        
    filecolor=magenta,      
    urlcolor=blue           
}
\usepackage[normalem]{ulem}

\newcommand{\equref}[1]{Eq.~(\ref{#1})}

\newcommand{\figref}[1]{Fig.~\ref{#1}}

\newcommand{\tableref}[1]{Table~\ref{#1}}

\renewcommand{\approx}{\simeq}

\renewcommand{\vec}[1]{\boldsymbol{#1}}

\begin{document}
\title{Superconducting ground state study of Cr-based equiatomic high-entropy alloy through \texorpdfstring{\(\mu\)SR}{muSR}}

\author{Sonika Jangid}
\affiliation{Department of Physics, Indian Institute of Science Education and Research Bhopal, Bhopal, 462066, India}

\author{Rhea Stewart}
\affiliation{ISIS Facility, STFC Rutherford Appleton Laboratory, Didcot OX11 0QX, United Kingdom}

\author{Adrian D. Hillier}
\affiliation{ISIS Facility, STFC Rutherford Appleton Laboratory, Didcot OX11 0QX, United Kingdom}

\author{R.~P.~Singh}
\email[]{rpsingh@iiserb.ac.in}
\affiliation{Department of Physics, Indian Institute of Science Education and Research Bhopal, Bhopal, 462066, India}

\begin{abstract}
High-entropy alloy superconductors, characterized by extreme chemical disorder and complex electronic environments, have attracted significant attention as model systems for exploring superconductivity in disordered materials. Here, we investigate a Cr-based equiatomic HEA, Cr-V-Ti-Nb-Ta, which contains a magnetic 3d element, providing an opportunity to examine the influence of magnetic elements on superconductivity in highly disordered systems. Despite expected magnetic pair-breaking, this alloy exhibits bulk type-II superconductivity with a transition temperature of $T_c = 2.33(3)$ K and a high upper critical field. Transverse-field $\mu$SR measurements reveal an s-wave superconducting gap close to the BCS value, while zero-field $\mu$SR suggests preserved time-reversal symmetry. These results establish Cr-V-Ti-Nb-Ta as a promising platform for exploring the interplay between disorder, magnetism and superconductivity in high entropy alloys.

\end{abstract}
\maketitle

\section{Introduction}
The discovery of superconductivity in high-entropy alloys (HEAs) has introduced a new dimension to the study of quantum materials, where extreme chemical disorder coexists with long-range electronic coherence \cite{kovzelj2014discovery}. In these materials, several principal elements (five or more) occupy a common crystallographic site in equiatomic or near equiatomic ratios, leading to strong potential fluctuations and high disorder, but often stabilizing simple crystalline phases and exhibiting exceptional mechanical properties \cite{yeh2004nanostructured, yeh2004formation, jien2006recent, cantor2004microstructural, zhang2008solid, otto2013relative, gludovatz2014fracture, huang2018twinning, zou2015ultrastrong, lee2007effect, tsai2013sluggish}. Their inherent chemical disorder often gives rise to remarkable superconducting properties, such as high upper critical fields, large critical current densities, and robust superconductivity under extreme pressure and significant disorder \cite{motla2023superconducting, jangid2024superconductivity, jangid2025lightweight, jung2022high, leung2022evidence, guo2017robust}. The vast compositional flexibility of HEAs enables precise tuning of crystal structure, disorder levels, and superconducting characteristics, thereby expanding their application landscape from structural and functional materials to superconducting devices and magnets \cite{marques2021review, pickering2021high, castro2021overview}.

To date, most studies on HEA superconductors have focused on optimizing $T_c$ through dependence on valence-electron concentration (VEC), with the majority of reported systems composed of non-magnetic 4d-5d elements \cite{kovzelj2014discovery, motla2023superconducting, jung2022high, leung2022evidence, guo2017robust}. However, introducing magnetic 3d elements adds a new degree of complexity, as localized magnetic moments can interact with the superconducting state. This interplay between magnetism and superconductivity remains one of the most persistent and fundamental challenges in condensed matter physics \cite{maple1995interplay}. Magnetic impurities are well known to suppress superconductivity through spin-flip scattering, a mechanism that disrupts Cooper pair formation and reduces the superconducting transition temperature $T_c$ \cite{rosa2014possible, hsieh2018cr, ding2022effect}. This phenomenon has been well established in elemental and intermetallic superconductors and has been explored in unconventional systems such as cuprates and iron pnictides \cite{ding2022effect, li2016chemical, bazargan2007localization, hsieh2018cr}. However, implications for strongly disordered superconducting systems, such as high-entropy alloys, remain largely unexplored. HEA superconductors, particularly those incorporating magnetic 3d transition metals such as Cr, offer a unique platform for investigating this issue \cite{liu2021superconductivity}.  Unlike conventional dilute impurity systems, Cr-based HEAs intrinsically incorporate 3d magnetic transition metal elements into a chemically disordered matrix, allowing the study of magnetic pair-breaking effects in a regime where both structural and electronic disorder is inherent \cite{liu2021superconductivity}. These effects can sometimes lead to unconventional pairing states \cite{rosa2014possible, chen2018progress}. Interestingly, the inclusion of the Cr element in HEA was originally motivated by the search for improved corrosion resistance \cite{liang2025new}. As such, Cr-based HEA superconductors serve not only as robust structural materials but also as a fertile ground for examining the coexistence and competition of magnetism and superconductivity in disordered systems. 

This work aims to leverage the intrinsic disorder and magnetic content of the Cr-based equiatomic HEA Cr-Ti-V-Nb-Ta to explore the mechanisms of magnetic pair-breaking and their broader implications on superconductivity using bulk and microscopic tools such as muon spin relaxation and rotation ($\mu$SR) measurements. Our comprehensive measurements reveals that it crystallizes in a body-centered cubic (bcc) structure and exhibits bulk type-II superconductivity. Transverse field (TF) $\mu$SR measurements confirm fully-gapped, isotropic s-wave superconductivity in the weak coupling limit. Zero field (ZF) $\mu$SR shows that the time-reversal symmetry remains preserved in superconducting ground state.
\begin{figure*}[ht]
\begin{center}
\includegraphics[width=2.0\columnwidth]{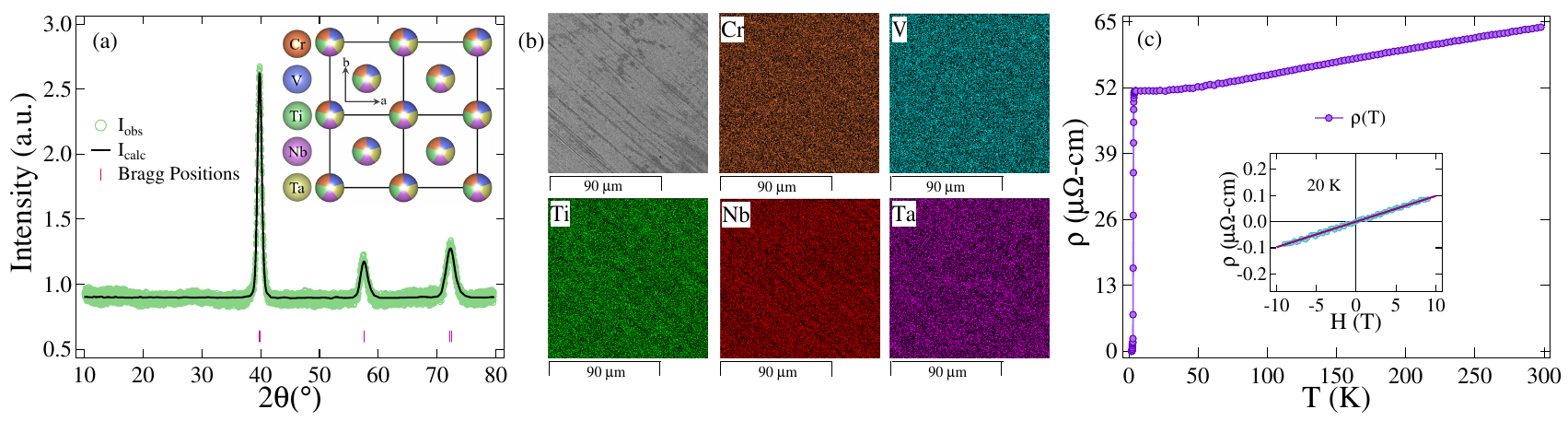}
\caption {\label{Fig1} (a) Room temperature x-ray diffraction pattern of Cr-V-Ti-Nb-Ta with the inset illustrating the atomic arrangement of constituent elements within its crystal structure. (b) EDS mapping of elements Cr, V, Ti, Nb, and Ta. (c) Temperature-dependent electrical resistivity of Cr-V-Ti-Nb-Ta measured from 1.9 K to room temperature in zero applied magnetic field. The inset presents the Hall resistivity measured at 20 K.}
\end{center}
\end{figure*}
\section{Experimental Details}
The polycrystalline sample of nominal composition  Cr$_{0.20}$V$_{0.20}$Ti$_{0.20}$Nb$_{0.20}$Ta$_{0.20}$ was synthesized by arc melting the high-purity elements Cr, V, Ti, Nb, and Ta in their stoichiometric ratio in an argon environment. Before alloy synthesis, a titanium getter was melted to eliminate residual oxygen. To promote chemical homogeneity and uniform phase formation, we repeatedly remelted and inverted the sample multiple times.

The crystal structure and phase purity were confirmed by powder x-ray diffraction using a PANalytic X$^{'}$Pert diffractometer with CuK$_{\alpha}$ ($\lambda = 1.5406 \text{\AA}$) radiation. Energy-dispersive x-ray spectroscopy (EDS) was performed using a scanning electron microscope (SEM) to examine the phase composition and homogeneity. Magnetization measurements were performed using a Quantum Design Magnetic Property Measurement System (MPMS-3) with a vibrating sample magnetometer. Electrical transport and specific heat measurements were performed using a Quantum Design Physical Property Measurement System (PPMS). Resistivity measurements were performed on a rectangular sample using the four-probe method, while heat capacity measurements were performed using a two-tau time-relaxation technique. Muon spin rotation and relaxation experiments were carried out on the MuSR instrument in transverse and longitudinal-field geometries, employing spin-polarized muons produced at the ISIS Neutron and Muon Source, Rutherford Appleton Laboratory (UK) \cite{hillier2019muons}.

\section{Results and Discussion}

\subsection{Sample Characterization}

X-ray diffraction pattern collected at room temperature is represented in \figref{Fig1}(a). The Le-Bail refinement was performed using HighScore Plus software, which confirms the crystallization in a bcc structure with space group Im3$\bar{m}$ (229) \cite{le1988ab}.
The broadening of observed peaks is attributed to a non-ideal sample preparation, probably a consequence of the inherent hardness of the material, as well as the significant structural disorder introduced by the presence of multiple constituent elements \cite{jangid2025superconducting}. The refined lattice parameters are $a=b=c=3.2021(2)$ \text{\AA} with a unit cell volume of $V=32.83(6)$ \text{\AA} and a calculated density $d= 8.58(6)$ g-cm$^{-3}$. In this structure, all constituent elements randomly occupy equivalent crystallographic sites, resulting in significant mixing of atomic sites, as illustrated in the inset of \figref{Fig1}(a). The elemental composition, determined by EDS at various points on the sample using SEM, yields an average composition of Cr$_{0.21}$V$_{0.20}$Ti$_{0.20}$Nb$_{0.19}$Ta$_{0.20}$, which closely matches the nominal composition within the experimental error. Elemental mappings via EDS, shown in \figref{Fig1}(b), demonstrate a uniform distribution of all elements, Cr, V, Ti, Nb, and Ta, across the sample, confirming phase homogeneity.
\begin{figure*}[ht]
\begin{center}
\includegraphics[width=2.0\columnwidth]{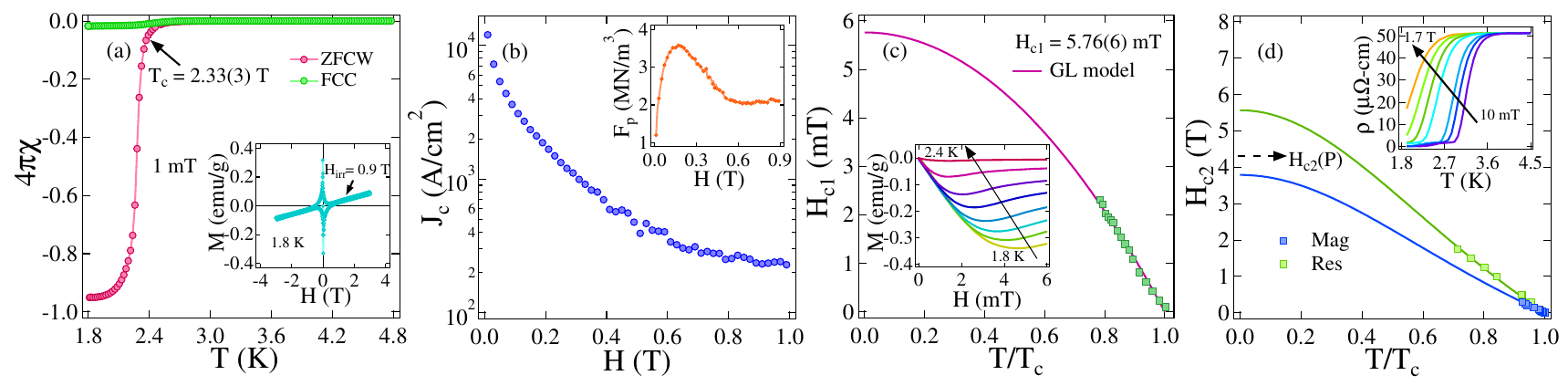}
\caption {\label{Fig2} (a) Temperature variation of DC magnetization for Cr-V-Ti-Nb-Ta measured in ZFCW and FCC modes, confirming the onset of superconductivity through a clear transition into the diamagnetic state. The inset shows the magnetization hysteresis loop at 1.8 K. (b) Magnetic field dependence of critical current density at 1.8 K, plotted on a semi-logarithmic scale and extracted from the MH loop. The inset shows the corresponding flux pinning force density at 1.8 K. (c) Temperature evolution of lower critical field, $H_{c1}$, for Cr-V-Ti-Nb-Ta fitted using GL model. The inset displays isothermal $M$-$H$ curves recorded between 1.8 K and 2.4 K. (d) Temperature evolution of upper critical field, $H_{c2}$, for Cr-V-Ti-Nb-Ta, also analyzed using the GL model. The inset displays resistivity curves under various applied magnetic fields in the vicinity of the superconducting transition.}
\end{center}
\end{figure*}

\subsection{Normal and Superconducting Properties}
Electrical resistivity data as a function of temperature for Cr-V-Ti-Nb-Ta, from 1.9 K to room temperature without any applied field, are shown in \figref{Fig1}(c). A sudden drop in resistivity at $T_c$ = 2.94(8) K confirms the emergence of superconductivity. The relatively low $T_c$ of Cr-V-Ti-Nb-Ta with respect to the non-Cr analog is consistent with the previous studies that increasing Cr concentration systematically suppresses $T_c$  and $H_{c2}$ in Cr-containing HEAs due to magnetic pair breaking effects associated with Cr incorporation \cite{liu2021superconductivity, xiao2023superconductivity, sarkar2022investigations}. In the normal state, the resistivity data exhibit a weak temperature dependence, which yields a residual resistivity ratio (RRR) $\frac{\rho_{300K}}{\rho_0}$ = 1.24(5). This close to unity RRR value suggests a poor metallic character and substantial atomic disorder, consistent with the inherent characteristics of high entropy alloys \cite{motla2021probing, motla2022boron, motla2022superconducting, motla2023superconducting, jangid2024superconductivity, jangid2025superconducting}.

To estimate carrier concentration, transverse resistivity $\rho_{xy}(H)$ was measured at 20 K. Data were fitted with a linear function [inset of \figref{Fig1}(c)], resulting in a Hall coefficient of $R_\mathrm{H} = 1.27(6)\times10^{-10}\ \Omega$-mT$^{-1}$. The positive sign of $R_\mathrm{H}$ indicates that holes are the major charge carriers. Applying the relation $R_\mathrm{H} = 1/ne$, the carrier concentration was calculated to be $4.92(1)\times10^{28}\ \mathrm{m}^{-3}$.

\vspace{1em}
\label{ch5:secMag}

DC magnetization as a function of the temperature for Cr-V-Ti-Nb-Ta, measured under the zero-field-cooled warming (ZFCW) and field-cooled cooling (FCC) protocols at an applied magnetic field of 1 mT, reveals a superconducting transition at 2.33(3) K, identified as $T_c$, as illustrated in \figref{Fig2}(a). The higher $T_c$ observed in resistivity compared to magnetization is due to possible inhomogeneous or filamentary superconductivity \cite{slebarski2014superconductivity, bianchi2001origin}. The significant separation between the ZFCW and FCC curves below $T_c$ indicates a strong flux pinning within Cr-V-Ti-Nb-Ta. Additionally, \figref{Fig2}(c) presents the magnetization hysteresis (MH) loop measured at 1.8 K, which exhibits a characteristic behavior of type-II superconductivity in Cr-V-Ti-Nb-Ta. The irreversibility field is estimated to be $H_{irr}=0.9$ T, marking the onset of irreversible vortex dynamics below which the magnetic flux lines remain pinned, giving rise to magnetization hysteresis. Above $H_{irr}$, thermal activation facilitates vortex motion, leading to reversible magnetization. The relatively large $H_{irr}$ is consistent with pronounced magnetization hysteresis and high critical current density observed in Cr-V-Ti-Nb-Ta. The critical current density is estimated using the MH loop measured at 1.8 K, based on Bean's critical state model, \cite{bean1962magnetization}:
\begin{equation}
    J_c = \frac{20\Delta M}{\left[a\left(1 - \frac{a}{3b}\right)\right]},
\end{equation}
where $\Delta M = M^{+} - M^{-}$ denotes the vertical width of the MH loop, and $a$ and $b$ represent the lateral dimensions of the sample with $a < b$. At 1.8 K in a magnetic field of 10 mT, the calculated critical current density is $J_c(1.8\ \mathrm{K},\ 10\ \mathrm{mT}) = 12001$ A-cm$^{-2}$, which is not only close to the commonly accepted threshold of $10^5$ A-cm$^{-2}$ for technological relevance \cite{jung2022high}, but is also comparable to the reported $J_c$ for as-cast Ta$_{1/6}$Nb$_{2/6}$Hf$_{1/6}$Zr$_{1/6}$Ti$_{1/6}$ HEA 
\cite{kim2020strongly}. The critical current density in a semi-logarithmic plot is represented in \figref{Fig2}(d). The field dependence of the flux pinning force density, defined as $\vec{F_p} = \vec{J_c} \times \vec{H}$, is shown in the inset of \figref{Fig2}(d) at 1.8 K. The maximum pinning force, $F_{p,\ \mathrm{max}} = 3.58$ MN-m$^{-3}$, is observed at an applied field of 180 mT.

Magnetization was measured as a function of the applied magnetic field at various temperatures between 1.8 K and 2.4 K [inset of \figref{Fig2}(b)] to determine the temperature dependence of the lower critical field, $H_{c1}(T)$. For each temperature, $H_{c1}$ was identified as the field at which the $M$-$H$ curve begins to deviate from its initial linear behavior. The temperature variation of $H_{c1}(T)$ follows the Ginzburg-Landau (GL) expression given by:
\begin{equation}
    H_{c1}(T)=H_{c1}(0)\left[1-\left(\frac{T}{T_{c}}\right)^{2}\right].
    \label{eqn2:Hc1} 
\end{equation}
The temperature evolution of the lower critical field, $H_{c1}(T)$, is presented in \figref{Fig2}(c). Data were fitted using \equref{eqn2:Hc1}, resulting in an extrapolated value of $H_{c1}(0)$ = 5.76(6) mT. Additionally, temperature-dependent magnetization measurements along with electrical resistivity versus temperature measurements [inset of \figref{Fig2}(d)], collected under different applied magnetic fields, were used to estimate the upper critical field [$H_{c2}(0)$]. A progressive reduction in $T_c$ is observed as the strength of the applied magnetic field increases. The determination of $H_{c2}(0)$ was based on this observed shift in $T_c$. The temperature dependence of $H_{c2}(T)$ was analyzed using the GL model, as presented in \figref{Fig2}(d),
\begin{equation}
H_{c2}(T) = H_{c2}(0)\left[\frac{(1-t^{2})}{(1+t^2)}\right].
\label{eqn3:Hc2}
\end{equation}
Here, $t = T/T_c$ represents the reduced temperature. By extrapolating the GL fit to absolute zero temperature, the upper critical field $H_{c2}(0)$ is estimated to be 3.79(8) T from magnetization and 5.57(8) T from resistivity measurements. The GL coherence length, $\xi_{\mathrm{GL}}(0)$, is related to the upper critical field by the relation:
\begin{equation}
    H_{c2}(0) = \frac{\Phi_0}{2\pi \xi^2_{\mathrm{GL}}(0)},
\end{equation}
where, $\Phi_0 = 2.07 \times 10^{-15}$ T-m$^2$ represents the magnetic flux quantum \cite{tinkham1996introduction}. Using the measured value of $H_{c2}(0)$, the GL coherence length at zero temperature is calculated as $\xi_{\mathrm{GL}}(0) = 93(2)$ \text{\AA}. This coherence length can then be used to determine the penetration depth using $H_{c1}(0)$ through the relation \cite{klimczuk2007physical}:
\begin{equation}
    H_{c1}(0) = \frac{\Phi_0}{4\pi \lambda^2_{\mathrm{GL}}(0)}\left(\ln{\frac{\lambda_{\mathrm{GL}}(0)}{\xi_{\mathrm{GL}}(0)}+0.12}\right).
    \label{eqn4:lambda_GL}
\end{equation}
At zero temperature, the GL penetration depth, $\lambda_{\mathrm{GL}}(0)$, is calculated to be 3238(103) \text{\AA}. The corresponding GL parameter, $\kappa_{\mathrm{GL}}$, calculated from the ratio $\lambda_{\mathrm{GL}}(0)/\xi_{\mathrm{GL}}(0)$, yields a value of 34(2), which is much higher than $1/\sqrt{2}$. This clearly classifies Cr-V-Ti-Nb-Ta as a type-II superconductor. The thermodynamic critical field, $H_c(0)$, is evaluated using the expression $H_{c1}(0)H_{c2}(0)=H_c^2(0)\ln\left({\kappa_{\mathrm{GL}}}\right)$ \cite{klimczuk2007physical}, based on the values of $H_{c1}(0)$, $H_{c2}(0)$, and $\kappa_{\mathrm{GL}}$, resulting in $H_c(0) = 78(7)$ mT.

Superconductivity in type-II materials can be destroyed by applying an external magnetic field exceeding the upper critical field. This suppression occurs primarily through two mechanisms: the orbital limiting effect and the Pauli paramagnetic effect. In the weak-coupling regime, the orbital limiting field in the Werthamer-Helfand-Hohenberg (WHH) model is expressed as \cite{werthamer1966temperature, helfand1966temperature}:
\begin{equation}
    H_{c2}(orb) = -\alpha T_c \left. \frac{dH_{c2}(T)}{dT}\right|_{T=T_{c}},
    \label{eqn5:Hc2_orb}
\end{equation}
\begin{figure}[ht]
\begin{center}
\includegraphics[width=1.0\columnwidth]{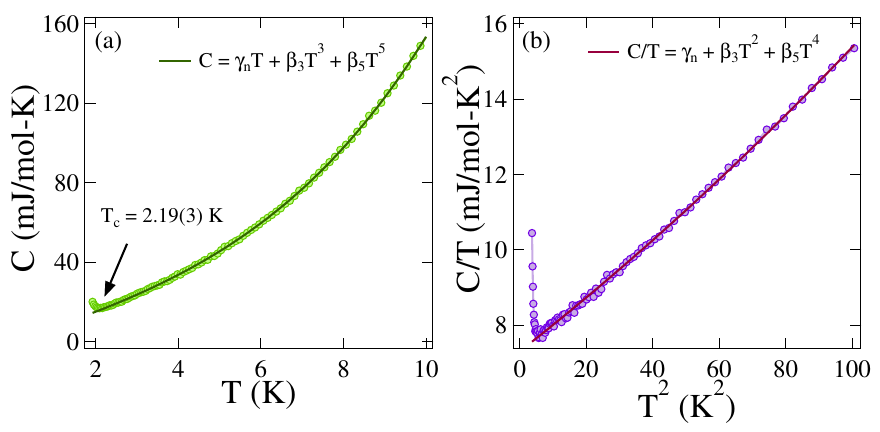}
\caption {\label{Fig3} (a) Temperature dependence of zero field heat capacity for Cr-V-Ti-Nb-Ta exhibiting a pronounced jump near the superconducting transition temperature. (b) $C/T$ vs $T^2$ curve under zero applied magnetic field. In the normal state region, the data are fitted using the Debye-Sommerfeld model.}
\end{center}
\end{figure}For a BCS dirty-limit superconductor, the constant $\alpha$ is taken as 0.693, using the slope $\left. T_c \frac{dH_{c2}(T)}{dT}\right|_{T=T_{c}} = -3.10(6)$ T, $H_{c2}({\mathrm{orb}})$, it is evaluated to be 2.15(6) T. In the weak-coupling limit of BCS theory, the Pauli limiting field is given by $H_{c2}(\mathrm{P}) = 1.86 T_c$ \cite{chandrasekhar1962note, clogston1962upper}. Using $T_c = 2.33(3)$ K, this yields a value of $H_{c2}({\mathrm{P}}) = 4.33(3)$ T. The Maki parameter, defined as $\alpha_\mathrm{M} = \sqrt{2} H_{c2}({\mathrm{orb}})/H_{c2}({\mathrm{P}})$, quantifies the relative contributions of the orbital and Pauli paramagnetic effect in Cooper pair breaking. For Cr-V-Ti-Nb-Ta, the Maki parameter is calculated to be 0.70(2) \cite{maki1966effect}. This value being less than unity indicates that the dominant pair-breaking mechanism is the orbital effect, with a negligible influence from the Pauli paramagnetic effect.

Defined as the ratio between the thermal energy ($k_\mathrm{B}T$) and the condensation energy within the coherence volume, the Ginzburg number ($G_i$) quantifies the relative strength of the thermal fluctuations and is expressed as \cite{blatter1994vortices}:
\begin{equation}
    G_i = \frac{1}{2}\left(\frac{k_\mathrm{B}\mu_0\tau T_c}{4\pi\xi^3_{\mathrm{GL}}(0)H^2_c(0)}\right)^2,
\end{equation}
where $k_\mathrm{B}$ is the Boltzmann constant and $\tau$ represents the anisotropy ratio, which is 1 for the cubic structure of Cr-V-Ti-Nb-Ta. Using the values $\xi_{\mathrm{GL}}(0) = 93(2)$ \AA, $H_c(0) = 78(7)$ mT, and $T_c = 2.33(3)$ K, the Ginzburg number was calculated as $G_i = 2.08(2) \times 10^{-7}$, which is comparable to the typical values observed in low-temperature superconductors ($\approx10^{-8}$) that indicate weak thermal fluctuations \cite{jangid2024superconductivity}.

\label{ch5:secHeat}
The heat capacity measurements for Cr-V-Ti-Nb-Ta under zero applied magnetic field, presented in \figref{Fig3}(a) and (b), provide insight into its thermal behavior. An anomaly at 2.19(3) K marks its superconducting transition, as evidenced by a substantial jump in heat capacity. In the normal state, the temperature dependence of the heat capacity $C(T)$ is well described by the Debye-Sommerfeld model:
\begin{equation}
    C(T) = \gamma_nT +\beta_3T^3 + \beta_5T^5.
\end{equation}
Here, $\gamma_n$ represents the Sommerfeld coefficient that accounts for the electronic contribution to the heat capacity, while $\beta_3$ and $\beta_5$ correspond to the phononic and anharmonic contributions, respectively. By extrapolating the fitted curve to absolute zero, the extracted parameters are $\gamma_n$ = 7.51(2) mJ-mol$^{-1}$K$^{-2}$, $\beta_3$ = 0.071(2) mJ-mol$^{-1}$K$^{-4}$, and $\beta_5$ = 0.008(2) mJ-mol$^{-1}$K$^{-6}$. The Debye temperature, $\theta_D$, is calculated using $\beta_3$ via the following relation \cite{kittel2005introduction}:
\begin{equation}
    \theta_\mathrm{D} = \left(\frac{12\pi^4RN}{5\beta_3}\right)^{\frac{1}{3}},
    \label{thetaD}
\end{equation}
where $R = 8.31$ J-mol$^{-1}$K$^{-1}$ denotes the universal gas constant, and $N$ represents the number of atoms per formula unit, taken as 1 for Cr-V-Ti-Nb-Ta. After substituting $R$, $N$, and $\beta_{3}$, \equref{thetaD} yields a Debye temperature $\theta_{\mathrm{D}}=301(9)$ K. The density of states at the Fermi level, $D_{C}(E_{\mathrm{F}})$, can be calculated using the formula for a non-interacting system:
\begin{equation}
    \gamma_n = \left(\frac{\pi^2k^2_\mathrm{B}}{3}\right)D_C(E_\mathrm{F}).
    \label{McMillan}
\end{equation}
By substituting the value of $\gamma_{n}$, $D_{C}(E_{\mathrm{F}})$ is evaluated as 3.18(1) states/eV-f.u. Furthermore, the electron-phonon coupling constant, $\lambda_{e-ph}$, which is a measure of the interaction strength between electrons and lattice vibrations, depends on both $\theta_{\mathrm{D}}$ and $T_{c}$ via the McMillan equation \cite{mcmillan1968transition}:
\begin{equation}
\lambda_{e-ph} = \frac{1.04+\mu^{*}\ln{\left(\theta_{\mathrm{D}}/1.45T_{c}\right)}}{\left(1-0.62\mu^{*}\right)\ln{\left(\theta_{\mathrm{D}}/1.45T_{c}\right)}-1.04}.
\label{eqn6:e-ph}
\end{equation}
Here, $\mu^{*}$ represents the screened Coulomb potential which is taken as 0.13 for intermetallic materials \cite{mcmillan1968transition}, and after inserting the values of $\theta_{\mathrm{D}}$ and $T_c$ into \equref{McMillan}, we obtain an electron-phonon coupling constant of $\lambda_{e-ph}$ = 0.525(6). This value places Cr-V-Ti-Nb-Ta in the weak-coupling regime. To examine the superconducting gap behavior for Cr-V-Ti-Nb-Ta, TF $\mu$SR measurements were employed.

\subsection{Muon Spin Relaxation and Rotation Measurement} 
\begin{figure*}[ht]
\begin{center}
\includegraphics[width=2.0\columnwidth]{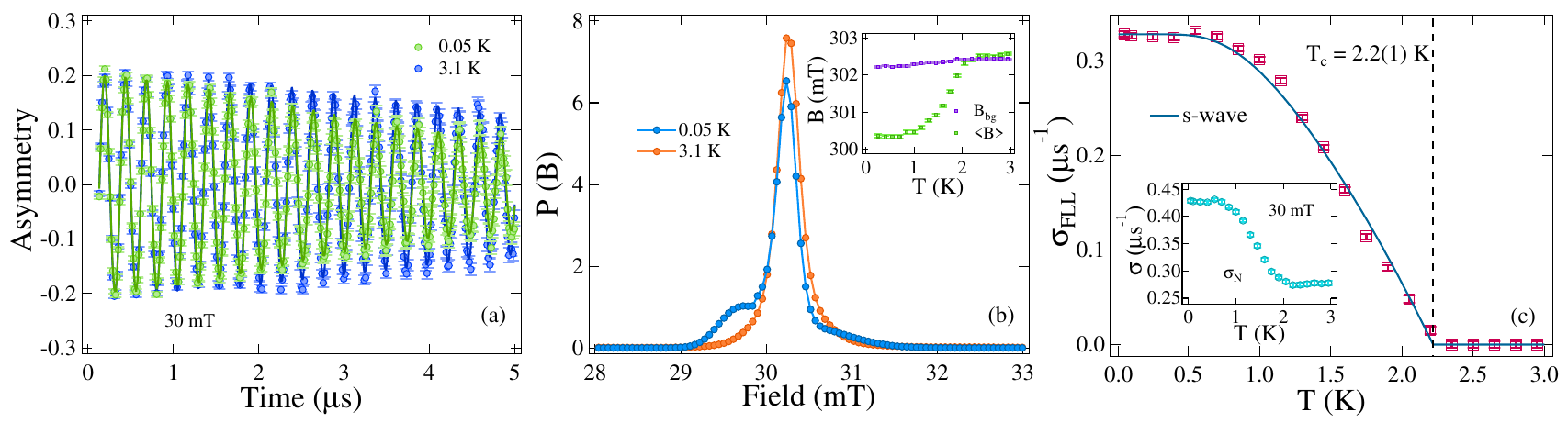}
\caption {\label{Fig4}(a) TF asymmetry spectra measured below (0.05 K) and above (3.1 K) the $T_c$ for Cr-V-Ti-Nb-Ta, recorded in FLL state at 30 mT applied field. The solid lines are fit to \equref{eqn9:TF_aymm}. (b) Probability distribution of magnetic field in superconducting (0.05 K) and normal (3.1 K) states calculated using the maximum entropy algorithm. The inset shows the internal magnetic field as a function of temperature, where green squares represent the field experienced by the sample while violet squares represent the background field. (c) Temperature dependence of the relaxation rate in the FLL state, fitted using the s-wave model. The inset shows the total relaxation rate as a function of temperature.}
\end{center}
\end{figure*}

To probe the superconducting ground state of Cr-V-Ti-Nb-Ta at the microscopic level, $\mu$SR experiments were conducted \cite{hillier2022muon}. The transverse field $\mu$SR measurements were performed under an applied magnetic field of 30 mT, oriented perpendicular to the initial direction of spin of the muon. This field strength lies between the lower and upper critical fields ($H_{c1} < H < H_{c2}$), allowing the formation of a flux-line lattice (FLL) within the mixed state of the superconductor. The sample was first cooled in the presence of the magnetic field, and data were collected as the sample warmed, following the field-cooled-warming protocol. The asymmetry spectra obtained above and below $T_c$ are shown in \figref{Fig4}(a). 
At 0.05 K, the asymmetry spectra show a more rapid relaxation compared to 3.1 K, which is attributed to the inhomogeneous magnetic field distribution arising from the formation of a flux line lattice. The observed reduction in depolarization above T$_c$ (3.1 K) is due to randomly oriented nuclear moments that remain static on the time scale of the muon. The corresponding probability field distributions in the vortex state at 0.05 K and the normal state at 3.1 K, obtained using the maximum entropy algorithm (MaxEnt), are presented in \figref{Fig4}(b). At 0.05 K ($T < T_c$), two distinct peaks are observed: one corresponds to the applied magnetic field detected by muons stopping in the sample holder, while the second reflects the internal field distribution caused by FLL. At $T = 3.1$ K, a single peak appears at the applied field value, indicating that Cr-V-Ti-Nb-Ta is in the normal state. The inset of \figref{Fig4}(b) presents the internal magnetic field distribution as a function of temperature. $B_{bg}$ denotes the temperature-independent background field, while $\left\langle B \right\rangle$ reflects the reduction in the applied field ($B_{app}$) due to magnetic flux expulsion as the material enters the superconducting state below $T_{c}$. Above $T_{c}$, $\left\langle B \right\rangle$ returns to $B_{app}$. The time-dependent asymmetry data were optimally described using a damped Gaussian oscillatory 
function, as given in \cite{maisuradze2009comparison, weber1993magnetic}.
\begin{equation}
\begin{split}
    A(t)& = A_1\exp\left(-\frac{1}{2}\sigma^2t^2\right)\cos(\gamma_\mu B_1t+\phi) \\
    &+ A_{bg}\cos(\gamma_\mu B_{bg}t+\phi).
\end{split}
\label{eqn9:TF_aymm}
\end{equation}
Here, $\phi$ denotes the initial phase, $A_{1}$, $B_{1}$, and $\sigma$ represent the asymmetry, mean-field, and relaxation rate of the sample, and $\gamma_{\mu}/2\pi = 135.5$ MHz/T is the gyromagnetic ratio of the muon. The parameters $A_{bg}$ and $B_{bg}$ correspond to the background asymmetry and the contributions from magnetic fields arising from muons stopping in the sample holder or the cryostat walls. In the normal state or above $T_c$, the value of $\sigma$ remains nearly constant, primarily due to the contribution of the nuclear dipolar field, denoted as $\sigma_{N}$. To isolate the superconducting
component $\sigma_{FLL}$, a value of $\sigma_{N} = 0.27(6)\ \mu \text{s}^{-1}$ was subtracted from the total $\sigma$ using the following relation.
\begin{equation}
\sigma_{FLL}^2 = \sigma^{2} - \sigma_{N}^{2}.
\label{eqn12:sigma}
\end{equation}
The temperature dependence of the calculated $\sigma_{FLL}$ is shown in \figref{Fig4}(c), while the inset of \figref{Fig4}(c) shows the total $\sigma$. At temperatures around one-fourth of the critical temperature ($T_{c}$/4), $\sigma_{FLL}$ becomes nearly constant, suggesting a fully gapped superconducting energy gap consistent with conventional pairing symmetry. This behavior excludes the presence of nodes in the gap structure.
In superconductors exhibiting a hexagonal Abrikosov flux-line lattice and a high upper critical field, the magnetic penetration depth $\lambda$ is related to the superconducting relaxation rate $\sigma_{FLL}$ as follows \cite{sonier2000musr, brandt1988flux}:
\begin{equation}
\frac{\sigma_{FLL}^2(T)}{\gamma_{\mu}^2} = \frac{0.00371\Phi_{0}^2}{\lambda^{4}(T)}.
\label{eqn13:sigmaH}
\end{equation}
Here, $\Phi_{0}$ denotes the magnetic flux quantum. From this analysis, the magnetic penetration depth at absolute zero is estimated to be $\lambda(0)$ = 5720(27) \text{\AA}. The larger penetration depth obtained from $\mu$SR than from magnetization can be attributed to the sensitivity of $\mu$SR to the local field distribution in the vortex state, which is affected by vortex pinning and disorder, thereby enhancing the effective $\lambda$. In contrast, magnetization probes the bulk response and is less sensitive to local inhomogeneities. Such discrepancies are commonly observed, particularly in polycrystalline superconductors \cite{sonier2000musr, fesenko1991analytical, motla2022superconducting, jangid2025superconducting}.

To further probe the superconducting gap symmetry, the temperature dependence of the relaxation rate, $\sigma_{FLL}(T)$, is analyzed using the theoretical model for an s-wave BCS superconductor in the dirty limit under the local London approximation, described by the following expression \cite{carrington2003magnetic}:
\begin{equation}
    \frac{\sigma_{FLL}(T)}{\sigma_{FLL}(0)}=\frac{\lambda^{-2}(T)}{\lambda^{-2}(0)}=\frac{\Delta(T)}{\Delta(0)}\tanh{\left(\frac{\Delta(0)}{2k_\mathrm{B} T}\right)},
    \label{eqn16:Mu}
\end{equation}
where the temperature dependence of the superconducting energy gap, $\Delta(T)$ is described by the BCS approximation:
\begin{equation}
    \Delta(T) = \Delta_{0} \tanh\left[1.82\left\{1.018\left(\frac{T_{c}}{T} - 1\right)\right\}^{0.51}\right].
\end{equation}
This function effectively captures the behavior of the superconducting relaxation rate, as shown by the solid blue curve in \figref{Fig4}(c), yielding an extrapolated superconducting gap value of $\Delta(0)/k_\mathrm{B}T_c = 1.78(4) $, which is close to the conventional BCS value of 1.76 confirming the presence of weakly coupled isotropic s-wave superconductivity in Cr-V-Ti-Nb-Ta. 


To investigate the possibility of time reversal symmetry breaking in Cr-V-Ti-Nb-Ta, zero-field $\mu$SR experiments were performed. \figref{Fig5} presents time-dependent asymmetry spectra recorded below (0.1 K) and above (3.5 K) $T_{c}$. The absence of oscillations in the spectra suggests a lack of long-range magnetic ordering in the system. However, an enhanced relaxation rate in the superconducting phase would suggest the emergence of spontaneous magnetization, which is indicative of time-reversal symmetry breaking \cite{ghosh2020recent}. When no electronic magnetic moments are present, muon depolarization arises exclusively from the influence of randomly oriented nuclear magnetic moments. At low temperatures, where muon diffusion is negligible, the asymmetry spectra in ZF-$\mu$SR are typically described using the Gaussian Kubo-Toyabe (KT) model \cite{hayano1979zero}.
\begin{equation}
G_{\mathrm{KT}}(t) = \frac{1}{3}+\frac{2}{3}(1-\Delta^{2}t^{2})\mathrm{exp}\left(\frac{-\Delta^{2}t^{2}}{2}\right).
\label{eqn15:ZF}
\end{equation} 
Here, $\Delta$ denotes the relaxation of the muon spin arising from randomly oriented static nuclear moments at the muon site. The zero-field spectra are best described by using the following function:
\begin{figure}[ht]
\begin{center}
\includegraphics[width=1.0\columnwidth]{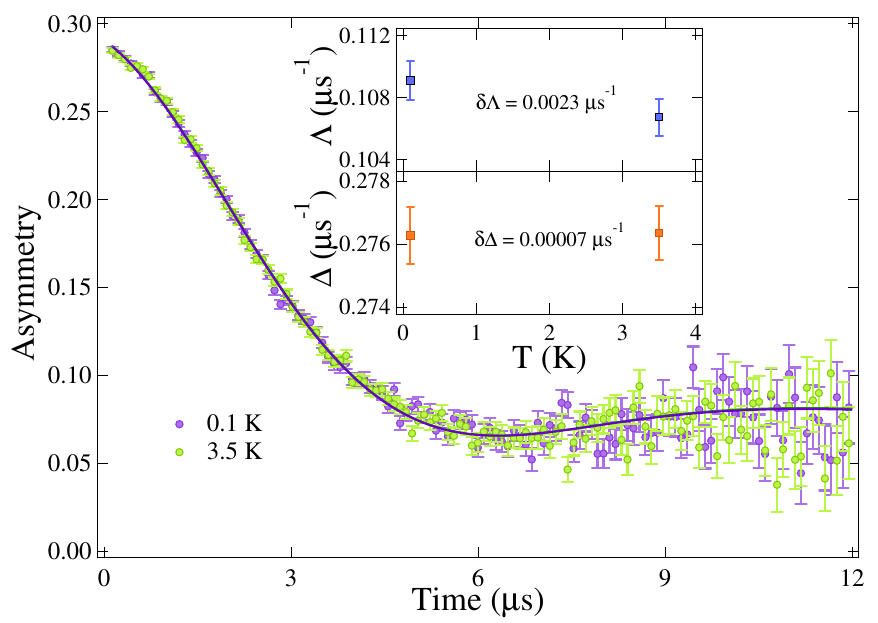}
\caption {\label{Fig5} Zero-field asymmetry spectra for Cr-V-Ti-Nb-Ta measured below (0.25 K) and above (5.0 K) $T_c$. The top panel of the inset shows the temperature dependence of relaxation rate $\Lambda$ while the bottom panel shows $\Delta$ with respect to temperature.}
\end{center}
\end{figure}
\begin{equation}
A(t) = A_{0}G_{\mathrm{KT}}(t)\mathrm{exp}(-\Lambda t)+A_{1}.
\label{eqn16:ZF_Re5.5Ta}
\end{equation}
Here, $A_{0}$ corresponds to the initial asymmetry related to the sample, $A_{1}$ represents the background asymmetry due to muons stopping in the sample holder, and $\Lambda$ corresponds to the relaxation rate associated with electronic contributions.
Within the sensitivity of the ZF-$\mu$SR measurements, the fitting parameters exhibit negligible change ($\Lambda=0.0023\ \mu \text{s}^{-1}$ and $\Delta=0.00007\ \mu \text{s}^{-1}$) across the superconducting transition as shown in the inset of \figref{Fig5}. These results demonstrate the absence of spontaneous magnetization, indicating that time-reversal symmetry remains preserved in Cr-V-Ti-Nb-Ta in the detection limit of MuSR.


\subsection{Electronic properties and Uemura plot}
Some key electronic parameters are calculated using experimentally determined values such as the Sommerfeld coefficient ($\gamma_n$), carrier concentration ($n$), and residual resistivity ($\rho_0$), using a set of established theoretical equations to complement the experimental findings for Cr-V-Ti-Nb-Ta. In particular, the Sommerfeld coefficient is connected to the effective mass ($m^*$) and carrier concentration ($n$) of the quasiparticles via the relation 
\begin{equation}
    \gamma_{n} = \left(\frac{\pi}{3}\right)^{2/3}\frac{k_\mathrm{B}^{2}m^{*}n^{1/3}}{\hbar^{2}}.
\end{equation}
Here, $\hbar=1.05\times10^{-34}$ J-s is the reduced Planck constant. Upon replacing $\gamma_n=7.51(3)$ mJ-mol$^{-1}$K$^{-2}$ and $n=4.92(1)\times10^{28}\ \mathrm{m}^{-3}$ (from Hall measurement), the effective mass $m^*=12.85(8)\ m_e$ is obtained. Fermi velocity $v_\mathrm{F}$ is related to $m^*$ and $n$ through the expression:
 \begin{equation}
     n=\frac{1}{3\pi^2}\left(\frac{m^*v_\mathrm{F}}{\hbar}\right)^3.
 \end{equation}
After inserting the values of $m^*$ and $n$, $v_\mathrm{F}$ is calculated to be 1.02(1)$\times10^5$ ms$^{-1}$. The mean free path ($l$) can be calculated using $\rho_0$, $m^*$, and $v_\mathrm{F}$ using the formula,
\begin{equation}
    l = \frac{3\pi^2\hbar^3}{e^2\rho_0m^{*2}v^2_{\mathrm{F}}},
\end{equation}
which was determined to be $l=18.41(5)$ \text{\AA}, a notably small value that is in line with the values reported for other high-entropy alloy superconductors \cite{motla2021probing, motla2022superconducting, motla2023superconducting, jangid2024superconductivity}. This short mean free path reflects the substantial atomic disorder introduced by the five different elements in the Cr-V-Ti-Nb-Ta lattice. Within the BCS formalism, the coherence length $\xi_0$ is defined as:
\begin{equation}
    \xi_0=\frac{0.18\hbar v_{\mathrm{F}}}{k_\mathrm{B}T_c}.
\end{equation}
\begin{figure}[ht]
\begin{center}
\includegraphics[width=1.0\columnwidth]{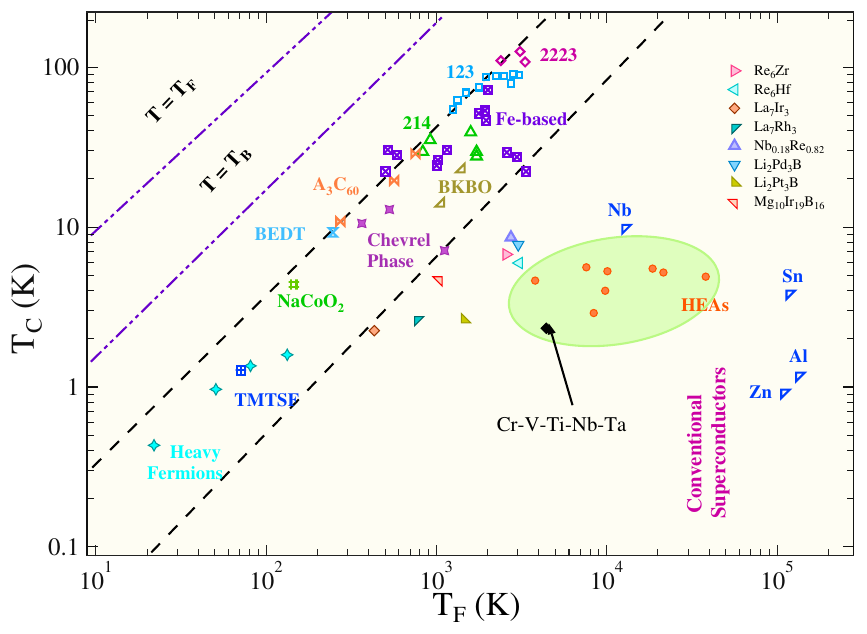}
\caption {\label{Fig6} The Uemura plot showing the relationship between $T_c$ and $T_\mathrm{F}$, where the region between the two dashed black lines represents the unconventional band, while Cr-V-Ti-Nb-Ta, indicated by a black square, lies outside this region.}
\end{center}
\end{figure}Upon replacing the values of $v_\mathrm{F}$ and $T_c$, $\xi_0$ is calculated to be 602(14) \AA. Superconductors are classified as clean or dirty according to the ratio of the BCS coherence length to the electronic mean free path, $\xi_0 / l$. In the case of Cr-V-Ti-Nb-Ta, $\xi_0/l=32(7) \gg l$, it is firmly placed in the regime of dirty-limits.

Uemura \textit{et al.} introduced a further classification of conventional and unconventional superconductors considering the ratio $T_c / T_{\mathrm{F}}$ \cite{uemura1988systematic, uemura1989universal, uemura1991basic}. Materials with $T_c / T_{\mathrm{F}}$ between 0.01 and 0.1 are referred to as unconventional, while those with $T_c / T_{\mathrm{F}} \geq 0.1$ fall into the conventional category. For a three-dimensional free-electron model, the Fermi temperature is expressed as \cite{hillier1997classification}:
\begin{table}[ht]
\caption{Parameters in the superconducting and normal state of Cr-V-Ti-Nb-Ta}
\label{Tab3}
\begin{center}
\setlength{\tabcolsep}{8 pt}
\begin{tabular}{c c c} 
\hline\hline
Parameters & Unit & Value  \\
\hline
VEC & & 5\\
$T_{c}$& K& 2.33(3)\\ 
$J_c$(1.8 K, 10 mT)& A-cm$^{-2}$&12001\\
$F_{p,\ max}$& MN-m$^{-3}$&3.58\\
$H_{c1}(0)$& mT& 5.76(6)\\                       
$H_{c2}^{mag, res}(0)$& T& 3.79(8), 5.57(8)\\
$H_{c2}(P)$& T&4.33(3)\\
$H_{c2}(orb)$& T& 2.15(6) \\
$\xi_{\mathrm{GL}}(0)$& \text{\AA}& 93(2)\\
$\lambda_{\mathrm{GL}}(0)^{mag,\ \mu}$& \text{\AA}& 3238(103), 5720(27)\\
$k_{\mathrm{GL}}$& &34(2)\\
$\gamma_{n}$&  mJ-mol$^{-1}$K$^{-2}$& 7.51(3) \\   
$\theta_\mathrm{D}$& K& 301(9)\\
$\lambda_{e-ph}$& &0.520(6)\\
${\left(\Delta(0)/k_\mathrm{B}T_c\right)^{\mu}}$& &1.78(4)\\
$\xi_{0}/l_{e}$& &  32(7)\\
$v_{\mathrm{F}}$& 10$^{5}$ ms$^{-1}$& 1.02(1)\\
$n$& 10$^{28}$m$^{-3}$& 4.92(1)\\
$T_{\mathrm{F}}$& K& 4418(50)\\
$m^{*}$/$m_{e}$&  & 12.85(8)\\
\hline\hline
\end{tabular}
\end{center}
\end{table}\begin{equation}
    k_\mathrm{B}T_\mathrm{F} = \frac{\hbar^2}{2m^*}\left(3\pi^2n\right)^{\frac{2}{3}}.
\end{equation}
Inserting the effective mass ($m^*$) and the carrier density ($n$) into the equation, we obtain a Fermi temperature $T_{\mathrm{F}} = 4418(50)$ K. The corresponding ratio $T_c/T_{\mathrm{F}} = 0.00052(1)$ places Cr-V-Ti-Nb-Ta well outside the unconventional superconductivity region (area between black dashed lines) and close to other HEAs shown in \figref{Fig6} \cite{jangid2024superconductivity, jangid2025lightweight, jangid2025superconducting, motla2021probing, motla2022boron, motla2022superconducting, motla2023superconducting}. A comprehensive list of both normal-state and superconducting parameters for Cr-V-Ti-Nb-Ta is provided in \tableref{Tab3}.

\section{CONCLUSION}
In summary, we have successfully synthesized a Cr-based equiatomic high-entropy alloy, Cr-V-Ti-Nb-Ta, which crystallizes in a body-centered cubic (bcc) structure. Comprehensive measurements confirm the emergence of bulk superconductivity with a critical temperature of 2.33(3) K. Magnetization studies clearly establish the material as a type-II superconductor. Remarkably, the upper critical fields derived from both magnetization and resistivity measurements are relatively high and approach the Pauli limit. This enhancement may be attributed to the strong disorder inherent to HEAs and spin-orbit interaction from heavy element Ta, warranting further experimental and theoretical investigation. Specific heat data indicate weak electron-phonon coupling, consistent with conventional superconductivity, while transverse-field $\mu$SR measurements support a fully gapped BCS s-wave pairing symmetry within the weak coupling regime. Furthermore, zero-field $\mu$SR measurements show no evidence of spontaneous internal fields, confirming that the time-reversal symmetry is preserved in the superconducting state. Overall, our results identify Cr-V-Ti-Nb-Ta as a valuable addition to the family of high-entropy alloy superconductors and offer new opportunities to investigate superconductivity in the presence of magnetic elements and strong chemical disorder.

\section*{ACKNOWLEDGEMENT} 
R.P.S. acknowledges the SERB, Government of India, for the Core Research Grant No. CRG/2023/000817. We acknowledge ISIS, STFC (U.K.), for the allocation of beamtime used for the $\mu$SR experiments.

\pagestyle{plain}
\addcontentsline{toc}{chapter}{Bibliography}

\bibliography{reference}

\end{document}